\documentclass[conference]{IEEEtran}
\IEEEoverridecommandlockouts
\usepackage{cite}
\usepackage{amsmath,amssymb,amsfonts}
\usepackage{graphicx}
\usepackage{textcomp}
\usepackage{xcolor}
\def\BibTeX{{\rm B\kern-.05em{\sc i\kern-.025em b}\kern-.08em
    T\kern-.1667em\lower.7ex\hbox{E}\kern-.125emX}}

\usepackage{eurosym}
\usepackage{array}
\usepackage{subfigure}
\usepackage{booktabs}
\usepackage{multirow}
\usepackage{dblfloatfix}

\usepackage{algorithm}
\usepackage[noend]{algpseudocode}

\algdef{SE}[DOWHILE]{Do}{doWhile}{\algorithmicdo}[1]{\algorithmicwhile\ #1}%

\usepackage{float}
\usepackage{psfrag}
\usepackage{hyperref}
\hypersetup{
    colorlinks,
    citecolor=blue,
    filecolor=blue,
    linkcolor=blue,
    urlcolor=blue,
    pdfstartview={XYZ null null 1.00}
}

\definecolor{gray1}{gray}{0.7}
\definecolor{gray2}{gray}{0.9}

\begin{document}

\title{Decentralized Task Allocation in Multi-Robot Systems via Bipartite Graph Matching Augmented with Fuzzy Clustering
\thanks{Copyright\copyright 2018 ASME. Personal use of this material is permitted. Permission from ASME must be obtained for all other uses, in any current or future media, including reprinting/republishing this material for advertising or promotional purposes, creating new collective works, for resale or redistribution to servers or lists, or reuse of any copyrighted component of this work in other works.}
}

\author{\IEEEauthorblockN{Payam Ghassemi\IEEEauthorrefmark{1}, and
Souma Chowdhury\IEEEauthorrefmark{2}}
\IEEEauthorblockA{\textit{Department of Mechanical and Aerospace Engineering} \\
\textit{University at Buffalo}\\
Buffalo, NY, 14260\\
Email: \IEEEauthorrefmark{1}payamgha@buffalo.edu,
\IEEEauthorrefmark{2}soumacho@buffalo.edu}}
%
%\author{\IEEEauthorblockN{Payam Ghassemi}
%\IEEEauthorblockA{\textit{Department of Mechanical \& Aerospace Engineering} \\
%\textit{University at Buffalo}\\
%Buffalo, NY, 14260\\
%payamgha@buffalo.edu}
%\and
%\IEEEauthorblockN{Souma Chowdhury}
%\IEEEauthorblockA{\textit{Department of Mechanical \& Aerospace Engineering} \\
%\textit{University at Buffalo}\\
%Buffalo, NY, 14260\\
%soumacho@buffalo.edu}
%}

\maketitle
%%%%%%%%%%%%%%%%%%%%%%%%%%%%%%%%%%%%%%%%%%%%%%%%%%%%%%%%%%%%%%%%%%%%%%
\begin{abstract}
Robotic systems, working together as a team, are becoming valuable players in different real-world applications, from disaster response to warehouse fulfillment services. Centralized solutions for coordinating multi-robot teams often suffer from poor scalability and vulnerability to communication disruptions. This paper develops a decentralized multi-agent task allocation (Dec-MATA) algorithm for multi-robot applications. The task planning problem is posed as a maximum-weighted matching of a bipartite graph, the solution of which using the blossom algorithm allows each robot to autonomously identify the optimal sequence of tasks it should undertake. The graph weights are determined based on a soft clustering process, which also plays a problem decomposition role seeking to reduce the complexity of the individual-agents' task assignment problems. To evaluate the new Dec-MATA algorithm, a series of case studies (of varying complexity) are performed, with tasks being distributed randomly over an observable 2D environment. A centralized approach, based on a state-of-the-art MILP formulation of the multi-Traveling Salesman problem is used for comparative analysis. While getting within 7-28\% of the optimal cost obtained by the centralized algorithm, the Dec-MATA algorithm is found to be 1-3 orders of magnitude faster and minimally sensitive to task-to-robot ratios, unlike the centralized algorithm.
\end{abstract}

%\vspace{-1.2cm}

%%%%%%%%%%%%%%%%%%%%%%%%%%%%%%%%%%%%%%%%%%%%%%%%%%%%%%%%%%%%%%%%%%%%%%
\section{INTRODUCTION} \label{sec:intro}
\subsection{Multi-agent Task Allocation}\label{ssec:mata-intro}
In contrast to the planning of actions for a single agent, Multi-Agent Task Allocation (MATA) is concerned with the allocation of tasks and resources among several agents that act together in the same environment to accomplish a common mission. In a MATA environment, there may be either conflicting or common goals pertaining to the different agents' decisions, which need to be coordinated~\cite{kuhn1992multi}. While centralized formulation and solution of MATA problems have traditionally dominated the fields of robotics, transportation, and IoT~\cite{korsah2013comprehensive, colistra2014problem}, decentralized methods have gained increasing prominence in recent years. This is partly due to increasing concerns regarding the scalability of purely centralized approaches and their vulnerability to communication disruptions, and partly driven by accelerated advancements in system/agent autonomy and Artificial Intelligence (AI) capabilities. In this paper, we consider task allocation for a team of robots in 2D space, and formulate the MATA problem as finding a set of optimal routes that minimize the overall costs incurred by the team ~\cite{kalmar2017multiagent}. Both a baseline centralized solution approach (based on multi-traveling salesman or mTSP problem) and a new decentralized framework (that combines bipartite graph theory and clustering concepts) are developed in this paper, and compared over a set of case studies involving varying numbers of robots and tasks. The remainder of this section briefly surveys the literature on centralized mTSP and auction-based (semi)-decentralized methods, and converges on the objectives of this paper.
%\vspace{1cm}

\subsection{Multi-Traveling Salesman Perspective of MATA}\label{ssec:lit}
The MATA problem, when defined by location-based tasks distributed over 2D space, is analogical to the Multi-Traveling Salesmen Problem (mTSP)~\cite{kalmar2017multiagent,bektas2006multiple} -- a generalization of the Traveling Salesman Problem (TSP)~\cite{gutin2006traveling,johnson1997traveling}. In mTSP, there are multiple salesmen or agents that need to visit a set of geographically distributed cities or locations, such that no city is visited more than once collectively by the agents. There are a number of variations on mTSP and we consider single depot mTSP, which is defined as~\cite{bektas2006multiple}: given a set of cities, one depot, and a cost metric, the objective is to find a set of routes for $m$ salesmen such that it minimizes the total cost of the $m$ routes. All salesmen are located in the depot and they return to the depot. In the literature, a number of approaches have been proposed for finding optimal solutions for the mTSP. Most of the approaches can be classified into two main classes~\cite{bektas2006multiple}: 1) algorithms based on Mixed-Integer Linear Programming (MILP) formulations, and 2) algorithms that transform the mTSP into the TSP. For solving the problem in each class, there are both exact solution and heuristic methods~\cite{bektas2006multiple}.

Although one of the earliest direct approaches for solving the mTSP was proposed by Laporte and Nobert~\cite{laporte1980cutting}, a branch-and-bound method proposed by Gavish and Srikanth~\cite{gavish1986optimal} can be considered to be one of the first attempt to solve a large-scale mTSP. Gavish and Srikanth \cite{gavish1986optimal} defined a Lagrangian problem for computing the lower bound of the main branch-and-bound algorithm. This method is limited to the symmetric mTSP. Gromicho et al.~\cite{gromicho1992exact} later solved the mTSP using a similar branch-and-bound algorithm but with a different approach to obtain the lower bound, namely a quasi-assignment relaxation instead of the Lagrangian formulation. This method solves the asymmetric mTSP with a fixed number of salesmen. There are also heuristic-based methods to solve the mTSP, such as by Zhang et al.~\cite{zhang1999team} and Ryan et al.~\cite{ryan1998reactive}, who respectively use Genetic Algorithms (GA) and Tabu search. Recently, Kalmar et al.~\cite{kalmar2017multiagent} proposed a modified GA method to solve the mTSP. These methods are slow to converge to optimal solutions, and often not suited for near-real-time applications (such as multi-robot real-time task-planning). 

The second class of the approaches, rather than solving the mTSP, investigate different strategies to transform the mTSP into a standard TSP. Gorenstein~\cite{gorenstein1970printing} solved the mTSP with $m$ salesmen as a standard TSP with $(m-1)$ additional home cities, where infinite costs are assigned to home-to-home distances and zero costs are assigned between the additional home cities and other cities. Jonker and Volgenant~\cite{jonker1988improved} improved the transformation of the mTSP to a standard TSP with a sparser edge configuration. This idea reduced the computational cost of the branch-and-bound scheme and created a less-degenerate TSP.

The aforementioned approaches are all centralized algorithms, typically characterized by exponentially growing computational effort \cite{nallusamy2009optimization}, as the number of agents and tasks increase. Moreover, if the MATA problem entails cost models that are highly nonlinear functions of the inter-city connecting edges, MILP formulations could become unreliable. Decentralized algorithms can, in principle, provide more tractable and dependable (if not provably optimal) solutions ~\cite{khamis2015multi}. Most of the existing decentralized solutions are consensus-based auction methods, where agents place bids on services or resources~\cite{shoham2008multiagent}. These approaches are not necessarily fully decentralized as they rely on a centralized auctioneer. In order to address this issue, Dai and Chen~\cite{dai2011multi} proposed a decentralized auction-based approach, where each agent acts as auctioneer and bidder to resolve any potential conflicts. In these approaches, multiple (thus time-consuming) biddings are necessary to make a decision, which also increases the inter-robot communication burden. However, consensus-based method do provide decision robustness under uncertainties, such as partial observance of mutual states within the robot team \cite{choi2009consensus}; further discussion of these characteristics is however not within the scope of this paper. A comprehensive review of methods on consensus-based decentralized auctions for task allocation can be found in \cite{choi2009consensus}. %Moreover, they do not guarantee to generate optimal solutions.

%\vspace{-0.2cm}
%\subsection{Assumptions}\label{ssec:assum}
\subsection{Objectives of this Paper}\label{ssec:obj}
Given the (above-described) characteristics of existing MATA methods, there remains an important opportunity for developing computationally-efficient and communication-frugal decentralized approaches to task allocation in multi-robot teams. While additional realistic considerations, such as partial observability and agent heterogeneity, can further complicate this problem, in this paper we focus on solving the efficient decentralized task planning problem subject to the following assumptions: i) all robots/agents are identical; ii) costs of all tasks can be evaluated deterministically; iii) there are no environmental uncertainties; and iv) each agent has full observation of all tasks and the state of other agents. Warehouse robotics~\cite{dAndrea2012guest}, such as Amazon's Kiva system~\cite{wurman2008coordinating}, can be considered to be a close match for an example, where the decentralized task allocation problem could be practically solved under these assumptions. 

While the long-term goal of this research is to develop stochastic learning-based algorithms that are capable of near-instantaneous decision-making and are insensitive to these assumptions, as a first step here we propose to develop a three-stage deterministic framework to decentralized planning. %(which could later on serve as the source for generating labels to train the learning based approaches). 
The first and second stages would seek to reduce the computational complexity of the problem being solved individually by each agent and identify a compact representation of the reduced MATA problem; and the third stage would allow each agent to identify optimal task sequences to undertake. The novel contribution of this paper lies in combining fuzzy clustering and bipartite graph matching to design this three-stage framework. Thus the objective of this paper is to develop this ``clustering--graph matching" based MATA framework, and investigate its performance in comparison to a state-of-the-art centralized mTSP implementation.  

%Our proposed algorithm can be integrated by the AI and reinforcement learning algorithms and can handle partially observability and uncertainty in the MATA. 
The remaining portion of the paper is organized as follows: The next section presents the problem definition and centralized mTSP formulation. Section~\ref{sec:decen} describes our proposed decentralized MATA framework. Results, encapsulating the performance of these methods on different-sized problems, are presented in Section~\ref{sec:results}. The paper ends with concluding remarks.

%This paper presents the development of a computational task planning algorithm supporting a decentralized swarm of robotic agents, with the backdrop of a warehouse management application. [Drawback and Prior in This Area]

%To the contrary, Kiva Systems LLC developed an agent-based autonomous pick-pack-and-ship warehouse system which does not rely on local decision making~\cite{wurman2008coordinating, enright2011optimization}.

% Summarize how the paper organized

%\subsection*{Multi-agent Planning} \label{ssec:intro-mas}
%\subsection*{Related Works} \label{ssec:intro-lit}

%%%%%%%%%%%%%%%%%%%%%%%%%%%%%%%%%%%%%%%%%%%%%%%%%%%%%%%%%%%%%%%%%%%%%%
\section{COLLABORATIVE MULTI-AGENT FORMULATION}\label{sec:probdef}
%Multi-agent Planning
\subsection{MATA: Defining Problem Components}
\noindent The multi-agent task allocation (MATA) problem is defined as the allocation of tasks and resources among several agents that act together without conflict in the same environment to accomplish a common mission. Each agent (robot) shares its state and its view of the world with other agents; in other words, the MATA is a tuple $\mathsf{T = <R, \{S_i\}, h, J, \{A_i\}, \{C_i\}, M, G>}$ where:
\begin{itemize}
\item[\textbullet] $\mathsf{R} = {1,\dots,m}$ is a finite non-empty set of agents.
\item[\textbullet] $\mathsf{S_i}$ is a set of state variables that represent the state of agent $i$ -- for example, its current location and battery state or traveled distance. Each agent can share their state variables with the other agents; i.e., full observability is assumed in the preliminary implementation presented in this paper.
\item[\textbullet] $\mathsf{h}$ is an integer value stating the maximum number of tasks that any robot can undertake.
\item[\textbullet] $\mathsf{J} = {1,\dots,n}$ is a finite non-empty set of active tasks that each robot is allowed to take.
\item[\textbullet] $\mathsf{A_{i}}: \mathsf{J}\times\mathsf{h}$ is a set of decisions of agent $i$ at iteration $k$, i.e., $\mathsf{A_{i}^k}$.
\item[\textbullet] $\mathsf{C_i}: \mathsf{A_{i}}\times\mathsf{J}$ is a finite set of state variables describing the relationship of each task $j$ with respect to agent $i$.
\item[\textbullet] $\mathsf{M}$ is a decision function that maps $\mathsf{S_i},\mathsf{C_i}$ to $\mathsf{A_{i}^k}$ at iteration $k$. 
\item[\textbullet] $\mathsf{G}$ is a metric that evaluates the total cost of the mission.
\end{itemize}
While the above definitions provide generic description of the problem components, in programmed practice most of these components are represented as pertinent vectors and matrices. The MATA problem is defined as finding the decision function $\mathsf{M}$ that generates the decision set $\mathsf{A_{i}^*}$, where $\mathsf{A_{i}^*}$ minimizes the total cost of the mission $\mathsf{G}$ and satisfies $\bigcup_{i=1}^m\mathsf{A_{i}^*}=\mathsf{J}$ and $\bigcap_{i=1}^m\mathsf{A_{i}^*}=\emptyset$. Next, we describe a modified mTSP formulation that will be serving as the benchmark centralized solution approach.
%The question is that how each agent takes a single decision so that is free of conflict and at the end each agent make a sequence of decisions that minimize the total cost of all agents.

\subsection{Centralized mTSP Formulation} \label{ssec:probcent}
\noindent The centralized MATA problem is formulated as a Mixed-Integer Linear Programming (MILP) problem~\cite{bektas2006multiple}, as given in Eq.~\eqref{eq:mTSPmilp} to \eqref{eq:mTSPmilp_SubTour}. The decision-space comprises a binary decision variable $z_{ij}$ and an integer variable $u_i$. The variable $z_{ij}$ becomes 1 if any robot takes task $j$ after finishing task $i$, and becomes $0$ otherwise. The variable $u_k$ shows the position of task $k$ in a sequence (tour). In Eq.~\eqref{eq:mTSPmilp} to \eqref{eq:mTSPmilp_SubTour}, $m$ and $h$ respectively represent the number of salesmen (robots) and the maximum number of tasks each robot can take in one tour; $c_{ij}$ is the cost metric for taking task $j$ after performing task $i$; it can essentially represent monetary cost, distance, time, or other cost function based on application.
\begin{align}
\label{eq:mTSPmilp}
\min_{z_{ij},u_i} f = \sum_{i=1}^n\sum_{j=1}^n c_{ij} z_{ij}
\end{align}
subject to
\begin{align}
\label{eq:mTSPmilp_consLeaveM}
\sum_{j=2}^n z_{1j} &= m\\
\label{eq:mTSPmilp_consEnterM}
\sum_{i=2}^n z_{i1} &= m\\
\label{eq:mTSPmilp_consEnter1}
\sum_{i=1}^n z_{ij} &= 1; \quad \forall~ j=2,\dots,n\\
\label{eq:mTSPmilp_consLeave1}
\sum_{j=1}^n z_{ij} &= 1; \quad \forall~ i=2,\dots,n\\
%\sum_{ij} x_{ij}t_{ij} &\leq T^s_{ij}; \quad \forall (i,j) \in A\\
\label{eq:mTSPmilp_SubTour}
u_i - u_j + h \cdot z_{ij} &\leq h-1; \quad \forall~ 2 \leq i \neq j \leq n
\end{align}
\noindent In the above set of equations, constraints~\eqref{eq:mTSPmilp_consLeaveM} and \eqref{eq:mTSPmilp_consEnterM} ensure that exactly $m$ agents depart from and return back to the depot (node 1). Constraints~\eqref{eq:mTSPmilp_consEnter1} and \eqref{eq:mTSPmilp_consLeave1} imply that each task is taken and finished by only one unique robot in the team. Constraint ~\eqref{eq:mTSPmilp_SubTour}, also known as Sub-tour Elimination Constraints (SCE), is utilized to remove sub-tours (circular closed routes not containing the depot). There are different mathematical formulations for the sub-tour elimination; in this work, we utilize one of the well-known formulations, called the Miller-Tucker-Zemlin constraints~\cite{miller1960integer}. Given this mTSP formulation, the overall cost incurred by the robot team can be estimated as:
\begin{align}
\label{eq:overallCost}
c_{\rm{total}} = \sum_{i=1}^n\sum_{j=1}^n c_{ij} z_{ij}
\end{align}
%and the task load of each agent $E_i$ is defined as:
%\begin{align}
%E_i = 
%\end{align}
%
%\subsection{Preliminaries}\label{ssec:prem}
%\begin{enumerate}
%\item Give a big picture about the actual problem.
%\item Introduce the assumptions to relax the problem.
%\end{enumerate}

%%%%%%%%%%%%%%%%%%%%%%%%%%%%%%%%%%%%%%%%%%%%%%%%%%%%%%%%%%%%%%%%%%%%%%
\section{DECENTRALIZED MATA ALGORITHM} \label{sec:decen} %DECENTRALIZED MULTI-AGENT TASK ALLOCATION
\subsection{Decentralized MATA: Overview}\label{ssec:overview}
\begin{figure}[h!]
\centering
\includegraphics[width=0.5\textwidth]{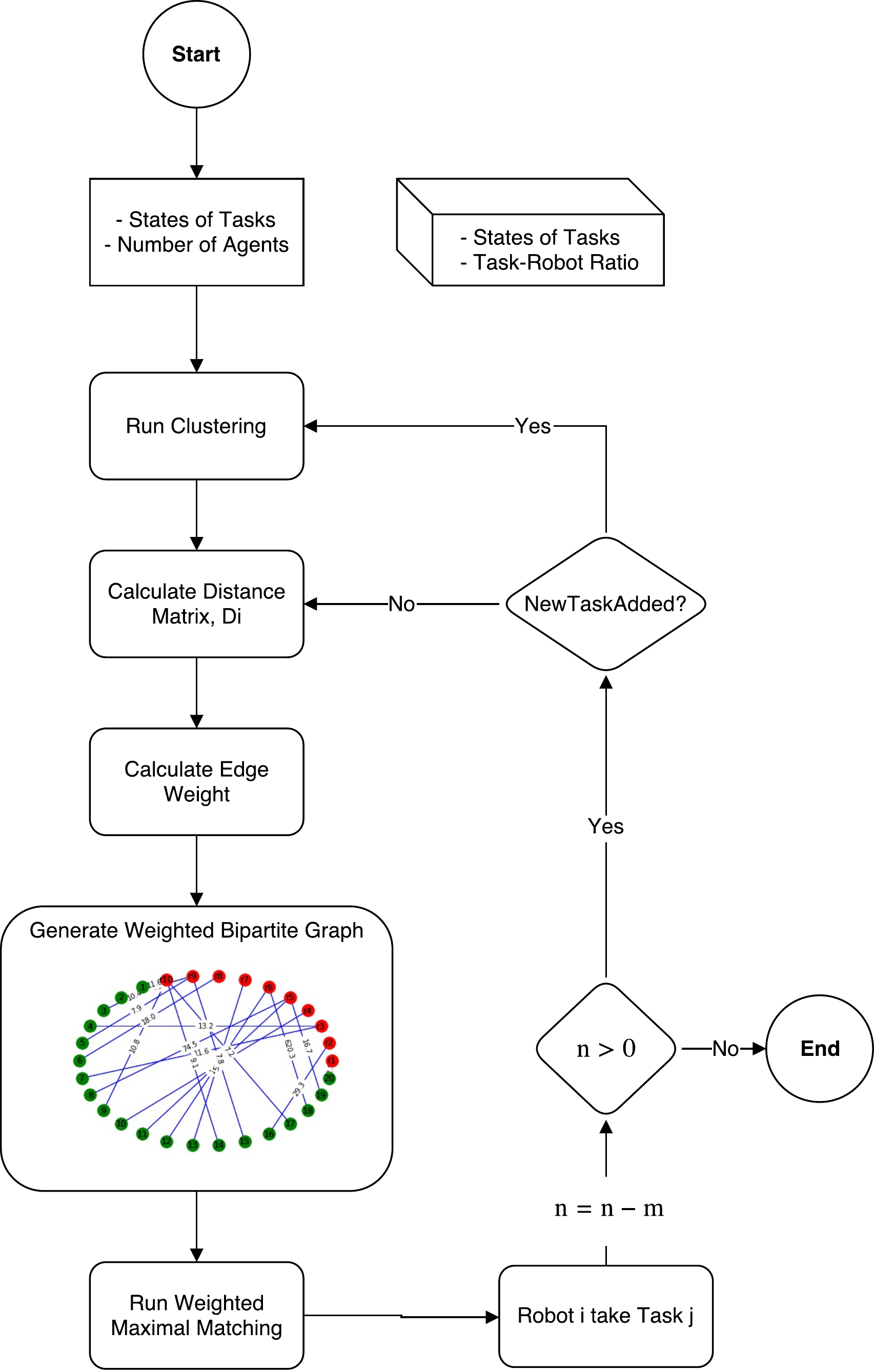}
\caption{Flowchart of the new decentralized MATA algorithm}
\label{fig:cse610hw1p3_DecMata}
\end{figure}

Figure~\ref{fig:cse610hw1p3_DecMata} depicts the flowchart of our proposed decentralized MATA or \textbf{Dec-MATA} algorithm. We pose the decentralized MATA as a maximum-weighted matching of a bipartite graph. Our MATA algorithm is composed of three components: 1) performing soft clustering to divide the tasks into groups (based on task location) to allow task-assignment problem decomposition; 2) use the clustering information and agents' states to transform the problem at each decision-making step into a weighted bipartite graph; and 3) solving a matching problem for the bipartite graph to assign a sequence of tasks to individual robots. Component 2 is generally the most challenging to accomplish. These three components are described next.
\begin{figure}[h!]
\centering
\includegraphics[width=0.4\textwidth]{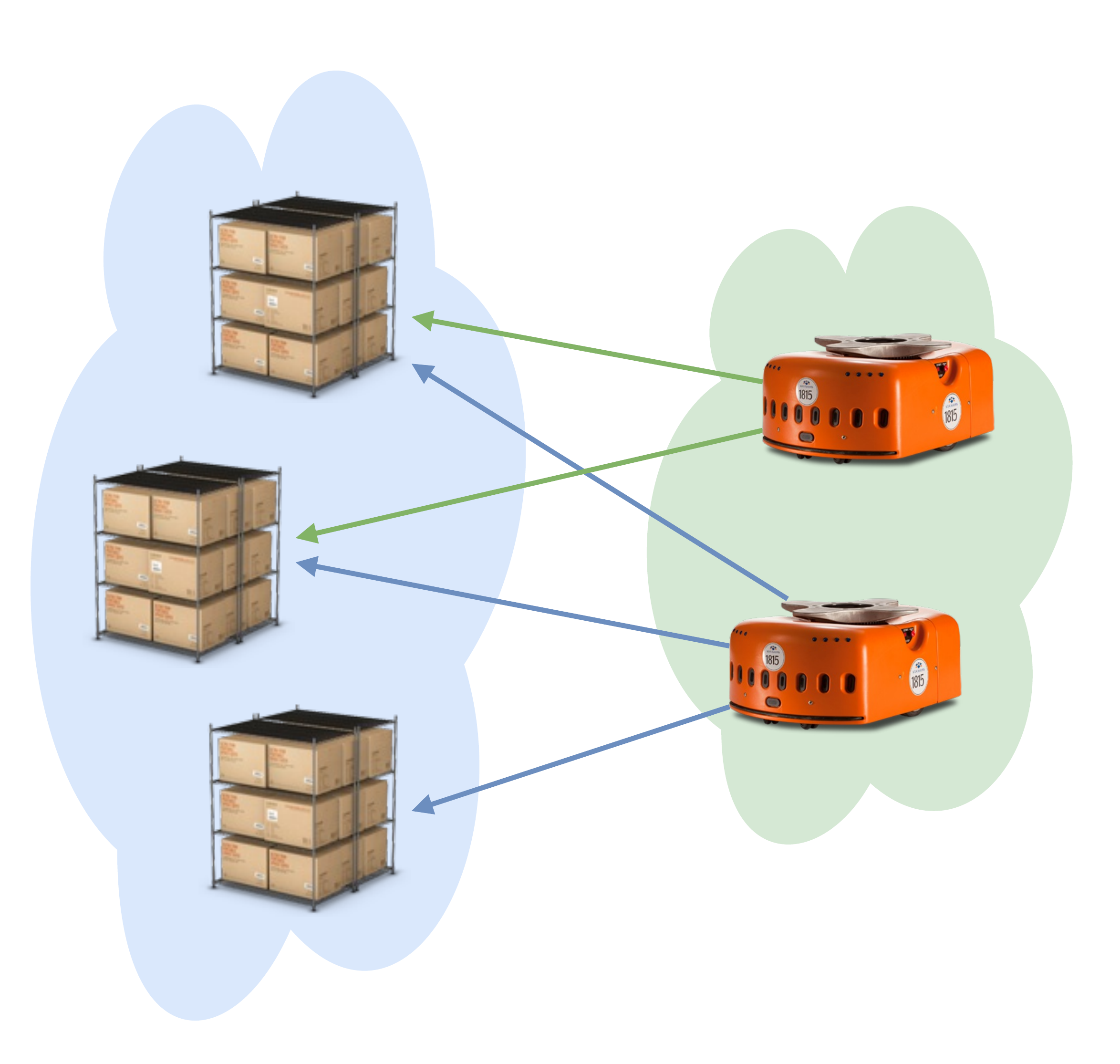} 
\caption{A representative weighted bipartite graph with 2 nodes in the robots set (Kiva robots) and 3 nodes in the tasks set (shelves).}
\label{fig:WBGraph}
\end{figure}

\subsection{Task Clustering}\label{ssec:task-clustering}
The purpose of clustering and graph construction is to identify and systematically represent the degree of relevance of groups of tasks to particular robots, given their current state -- thereby reducing the size of the task assignment problem to be solved by each robot. In this paper, relevance is judged based on the current location of tasks and robots. However, the underlying concept can be extended to apply to more generalized scenarios where the state space of the robots and the feature space of the tasks are more complicated. There is also an implicit assumption here that the number of tasks is always greater than the number of robots, and hence our aim is to identify task relevance not individually but in a group. In order to group the tasks according to their relative location, while allowing relevance to be defined in a smooth (as opposed to a binary, relevant/not relevant) manner, the Fuzzy C-Means clustering algorithm (FCM)~\cite{bezdek1984fcm} is used. FCM is a soft clustering method, which allows cluster overlap and is performed by solving the following minimization ~\cite{bezdek1981objective}:

\begin{align}
\label{eq:fcm}
b_{ij}^*\in \arg\min_{b_{ij}}\sum_{i=1}^m\sum_{j=1}^n b_{ij}^\gamma\|x_j - r_i\|
\end{align}
In Eq. \ref{eq:fcm}, $n$ and $m$ are the number of data points (in this case, number of tasks) and the number of clusters (in this case, number of robots), respectively; $\gamma>1$ is a user-defined coefficient that adjusts the degree of fuzzy overlap; $x_i$ is the $i$-th data point or task, and $r_i$ is the location of the $i$-th robot, serving as the associative (not necessarily exact geometric) center of the $i$-th cluster. Here, $b_{ij}$ is the degree of membership of $x_j$ in the $i$-th cluster and ranging from 0 to 1. For ease of implementation, in this paper, the number of fuzzy clusters is set equal to the number of robots. As shown in the algorithm flowchart (Fig. \ref{fig:cse610hw1p3_DecMata}), the clustering step needs to be repeated only if new tasks get added to the mission. 
%\noindent Using Eqs.~\eqref{eq:edgeWeight}-\eqref{eq:fcm}, the weighted bigraph is fully determined, in the following section we introduce a method to find the optimal tasks allocation.
%
\begin{figure*}[th!]
\centering
\begin{tabular}{cc}
\includegraphics[width=0.5\textwidth]{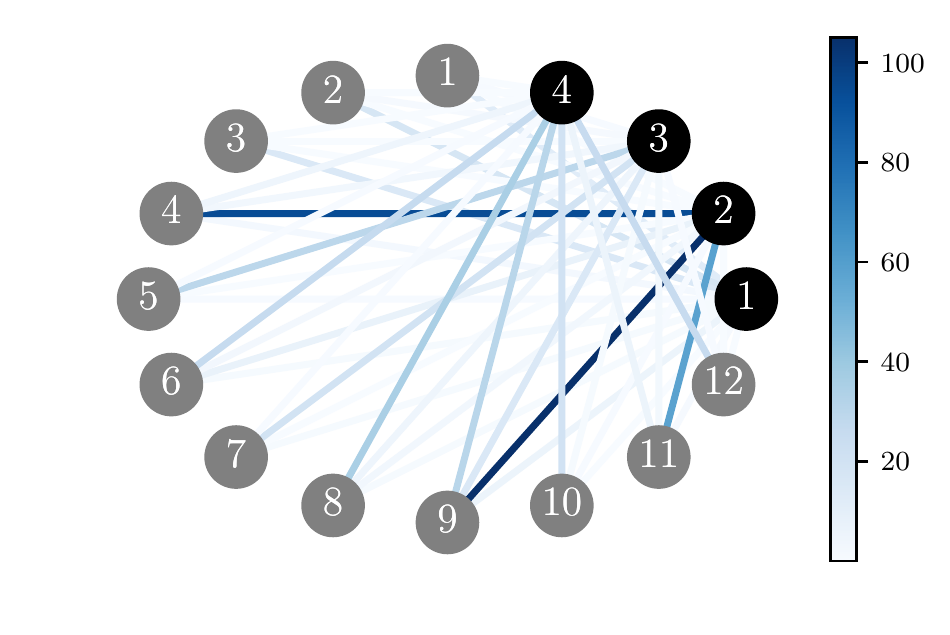} & 
\includegraphics[width=0.5\textwidth]{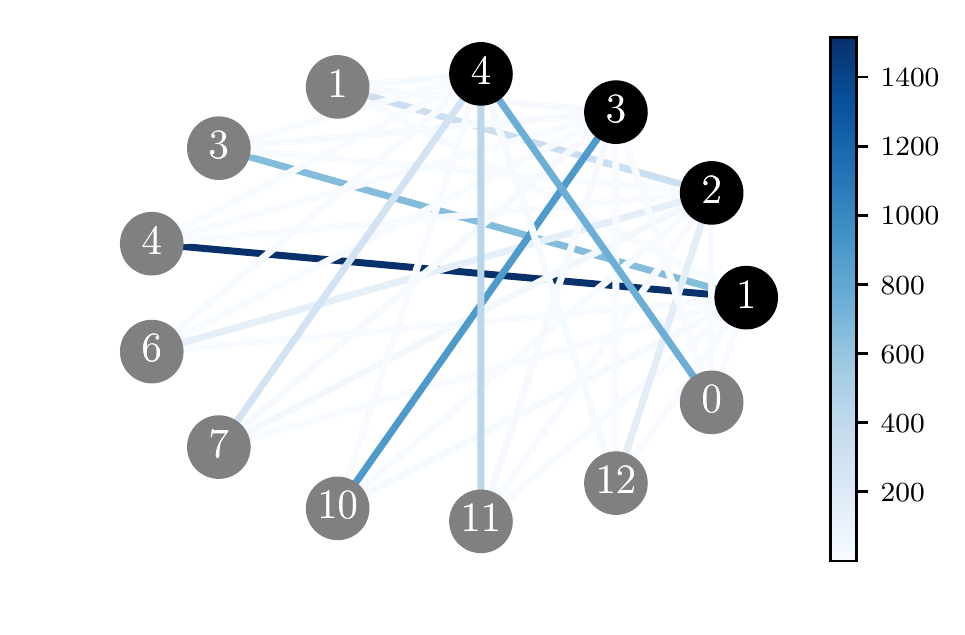}\\
(a) First decision-making iteration & (b) Second decision-making iteration   
\end{tabular}
\caption{Bigraph generated by the dec-MATA algorithm for a representative \textit{4 robots / 12 tasks} case (i.e., $m=4$, $n=12$): black and gray circles represent robots and tasks, respectively; the color of the edges represent the estimated weights of the connections between robots and tasks. \textit{The bigraph can be observed to vary from one decision-making iteration to another; since the robots collectively chose tasks 2, 5, 8 and 9 in the first iteration, these tasks are missing from the reduced bigraph constructed in the second iteration. However, the latter bigraph has one more task -- task 0 basically represents the depot, which become available for selection only after the first iteration; this is done to ensure that each robot undertakes at least 1 task.}  %Black and gray circles represent robots and tasks, respectively. Bigraph generated by the decentralized MATA algorithm for a representative \textit{4 robots -- 12 tasks} case (i.e., $m=4$ and $n=12$): the figures show how the bigraph varies from one decision-making iteration to another.
}
\label{fig:cse610hw1p3_wbg}
\end{figure*}
%
% % For next version of algorithm (W_th)
% \begin{figure*}[th!]
% \centering
% \begin{tabular}{cc}
% \includegraphics[width=0.5\textwidth]{Figures/mata_milp_28-May-2018_n10_p4_j20_Run1_graph1_Wth_0.pdf} & 
% \includegraphics[width=0.5\textwidth]{Figures/mata_milp_28-May-2018_n10_p4_j20_Run1_graph2_Wth_0.pdf}\\
% (a) First decision-making iteration & (b) Second decision-making iteration   
% \end{tabular}
% \caption{This figure illustrates how the bigraph changes after the first iteration during the application of Dec-MATA for the test case 4 (\textit{10 robots and 20 tasks}; i.e., $m=10$ and $n=20$). At the first iteration, }
% \label{fig:cse610hw1p3_wbg}
% \end{figure*}
% %
%
\begin{algorithm*}[th!]
\caption{Blossom algorithm (adopted from~\cite{slota2018graph})}\label{alg:blossom}
\begin{algorithmic}[1]
\Procedure{MatchGraph}{$G$}\Comment{Graph $G(V,E)$}
\State $M \gets \emptyset$ \Comment{Match $M$ initially empty}
\Do
\State $P \gets $ \textsc{BlossomAlg}($G,M$) \Comment{New augmenting path found with $M$, $G$}
\State $M \gets M\Delta P$ \Comment{Symmetric difference between $M$, $h$}
\doWhile{$P \neq \emptyset$}
\State \textbf{return} {$M$}
\EndProcedure

\Procedure{BlossomAlg}{$G, M$}\Comment{Graph $G(V,E)$ and matching $M(V_M, E_M)$}
\State $G^\prime \gets G$
\For{\textbf{all} $v\in V$}
	\State $marked(v) \gets$ \textbf{false}
\EndFor\label{alg:setting}
\For{\textbf{all} $v \in V$: $v \not\in V_M$, $marked(v)=$ \textbf{false}}
	\State $marked(v) \gets $ \textbf{true}
    \State $Q \gets v, Q_n \gets \emptyset$
    \State $T \gets v$ \Comment{Modified BFS tree}
    \While{$Q \neq \emptyset$}
    \For{\textbf{all} $u \in Q$}
    \For{\textbf{all} $w \in N^\prime(u)$}
    \If{$w \not\in V_M$}\Comment{$w$ is unsaturated}
    \State \Return{shortestPath($T,u,v$)+($u,w$)}
    \ElsIf{$marked(w) =$ \textbf{false}}\Comment{Not in $T$ yet}
    \State $x \gets y \in N(w): (y,w)\in E_M$ \Comment{Vertex $w$'s match}
    \State $marked(w), marked(x) \gets $ \textbf{true}
    \State $T \gets T + (u,w) + (w,x)$
    \State $Q_n \gets x$
    \ElsIf{abs(level($T,u$)-level($T,w$)) mod $2$ = even}\Comment{ٍEven level difference on non-tree edge means blossom found}
    \State $B \gets $ shortestPath($T,u,v$)+($u,w$)
    \State $G^\prime \gets G^\prime \cdot B$ \Comment{Contraction of blossom}
    \State $T \gets T \cdot B$
    \State $b \gets $ contracted blossom vertex
    \State $marked(b) \gets $ \textbf{true}
    \State $Q_n \gets b$
    \EndIf
    \EndFor
    \EndFor
    \EndWhile
\EndFor\label{alg:cal}
\textbf{return} $\emptyset$
\EndProcedure
\end{algorithmic}
\end{algorithm*}
\subsection{Weighted Bipartite Graph Transformation}\label{ssec:proWBG}
%Our problem, as defined in the Section~\ref{sec:probdef}, is quite similar to a weighted bipartite matching problem~\cite{schwartz2005fast} in the graph theory. 
In order to represent and analyze the task-robot relations, we use the concept of bipartite graphs, so-called bigraphs (popularly used in recommender systems~\cite{huang2007analyzing} and social network analysis~\cite{zhu2015measuring}). A bigraph is a graph whose vertices can be divided into two sets of vertices such that no two vertices in the same set are joined by an edge~\cite{asratian1998bipartite}. In this paper, we define our problem as a weighted bigraph $(\mathsf{R, J, E})$ during each decision time-period, where $\mathsf{R}$ and $\mathsf{J}$ are a set of robots (agents) and a set of tasks, respectively; and $\mathsf{E}$ represent a set of weighted edges that connect robots to available tasks, as shown in Fig.~\ref{fig:WBGraph}. This current bipartite graph definition (allowing task-robot one-to-one mapping) is applicable to MATA problems, where multiple agent collaboration on any single task is not required (rather prohibited).

In order to fully construct the representative weighted bipartite graph, we should determine the weights of edges, a typically challenging endeavor given the lack of any standard recommendations to this end. Here, the edge weights can be perceived as the strength of the potential relationship between the corresponding task and robot, and thus indicative of the degree of relevance of task $j$ to robot $i$. The weight, $w_{ij}$, of the bigraph edge $ij$ is defined here in terms of the states of robot $i$ and task $j$ as:
%\vspace{-0.2cm}
%
\begin{align}
\label{eq:edgeWeight}
{w}_{ij} = K_b b^*_{ij} \frac{c_{\rm{c},i}+1}{(c_{ij}+1)^2}
\end{align}
%
%\vspace{-0.2cm}
Here, the parameter $b^*_{ij}$, which relates task-$j$ to agent-$i$, is obtained from fuzzy clustering (Eq. \ref{eq:fcm}). The robots are assumed to start from a single depot (e.g., co-located charging stations). Hence, initial task assignment (considered as the zeroth decision-making iteration) is randomized. However, then onwards, each robot assumes the location of the current task (and hence membership of its cluster) that it is undertaking, allowing us to use the $b^*_{ij}$ relation (which previously represented task-cluster relatedness). Here, $K_b$ is a coefficient to adjust $b^*_{ij}$ with respect to the magnitude of $c_{ij}$, where $c_{ij}$ shows the cost of doing task $j$ (next) for robot $i$. The parameter $c_{\rm{c},i}$ is the cumulative cost incurred by robot $i$ up to that decision point. Note that the bigraph reduces by $m$ tasks, as tasks get done by the $m$ robots in each iteration. The depot (denoted as task 0) becomes available for selection iteration 2 onwards. Figure~\ref{fig:cse610hw1p3_wbg} depicts the weighted bigraphs generated at the first and second decision-making iteration for a representative\textit{ 4 robot -- 12 task} case. %Here, $w_{th}$ is a user-defined control parameter that imposes a threshold on generating an edge. In other words, if the weight of an edge connecting robot-$i$ and task-$j$ in the graph is lower than the threshold, they are assumed to have no relation. 
%In literature, there are various methods for perfect matching ($|R| = |J|$), where describes a way of simultaneously satisfying all robots and taking all tasks. In our problem, the number of tasks and agents are not the same, so the perfect matching is not possible.
%\subsection{Edge Weighting} \label{ssec:proEW}
%\subsection{Edge Filtering} \label{ssec:proEW}
%\subsection{Clustering} \label{ssec:proClust}

\subsection{Maximum Weighted Matching}
Once the weighted bipartite graph has been constructed, the final step (Fig.~\ref{fig:cse610hw1p3_DecMata}) is to solve the task assignment or allocation problem as a matching problem. This problem is defined as drawing a set of edges such that they do not share any vertices~\cite{lovasz2009matching}. A \textit{weighted maximum matching} method is used here to determine the optimal task assignment. Note that, until this point, all robots have taken the same (redundant) computational steps; however, this final step differs across robots, since each robot operates only on the portion of the bipartite graph relevant to itself (e.g., not consider tasks for which its bigraph connecting edge has a weight of zero). To perform this final step, an improved maximum matching algorithm proposed by Galil~\cite{galil1986efficient} is adopted. This algorithm is based on the classical blossom algorithm introduced by Edmonds~\cite{edmonds1965maximum}, which is known to run in polynomial time. The pseudocode of this algorithm is given in Algorithm~\ref{alg:blossom}. This algorithm produces the optimal decision function, $M$ (i.e. the task assignment set). It is important to note that the outcomes of this decentralized decision-making process are deterministic and synchronized (assuming a deterministic environment with perfect localization). Consequently, each robot will arrive at the same conflict-free optimal task assignment outcomes. Although, currently, a degree of redundancy of taking the same decision on-board each robot (w.r.t. the same task cluster) is allowed, more efficient implementations can be explored in the future. These future advancements could also help promote asynchronous multi-agent decision-making, which is a highly-challenging and an open area of research in itself~\cite{zedadra2017multi}.
%
%\subsection{Agent Elimination}
%\begin{figure*}[h!]
%\centering
%\begin{tabular}{c}
%\includegraphics[width=0.8\textwidth]{Figures/MAS_Experiment_Flowchart_Dataset.png}\\
%\includegraphics[width=0.8\textwidth]{Figures/MAS_Experiment_Flowchart_Eval.png}\\
%\end{tabular}
%\caption{Degree distribution for a random network with different probability values, $h$, ranging from $0.01$ to $0.4$.}
%\label{fig:cse610hw1p3_dist}
%\end{figure*}
%%%%%%%%%%%%%%%%%%%%%%%%%%%%%%%%%%%%%%%%%%%%%%%%%%%%%%%%%%%%%%%%%%%%%%
\section{NUMERICAL EXPERIMENTS} \label{sec:results}
\subsection{Problem and Framework Settings}\label{ssec:cases}
We design and execute a set of numerical experiments to investigate the performance and scalability of the proposed decentralized MATA algorithm (Section \ref{sec:decen}), and compare it with the centralized mTSP solution (Section \ref{ssec:probcent}). For solving the MILP problem resulting from the centralized approach~\eqref{eq:mTSPmilp}, the``\textit{Gurobi V8.0}"~\cite{gurobi} is run on the MATLAB\textsuperscript{\tiny\textregistered} R2016b. A combination of MATLAB libraries (e.g., fuzzy c-means clustering), Python libraries and original codes are used to implement our decentralized MATA framework.
Specifically, the \textit{"Python"} 3.6.0 and the 64-bit distribution of \textit{"Anaconda"} 4.3.0 are used to implement the decentralized MATA framework, and the ``\textit{networkx}" library is used for graph analysis and visualization. The simulations are executed on a system with Intel\textsuperscript{\tiny\textregistered} i7-6820HQ 2.70 GHz 4 Cores (8 Logical Processors) CPU and 16 GB RAM. The Gurobi (MILP) solver exploits all 8 logical processors, while the Python implementation of Dec-MATA does not exploit all the cores. Here the parameters of the Gurobi's solver are set as: if the computation time is more than $7200$s, the absolute MIP optimality gap ("\texttt{MIPGapAbs}") equals $1e-10$, otherwise this gap, \texttt{MIPGapAbs}, is set at $6$. Research data and codes related to Dec-MATA implementation can be found at: {\scriptsize\url{http://adams.eng.buffalo.edu/algorithms/multi-robot-algorithms/}}.

\begin{figure*}[h!]
\centering
\begin{tabular}{cc}
\includegraphics[height=0.32\textwidth]{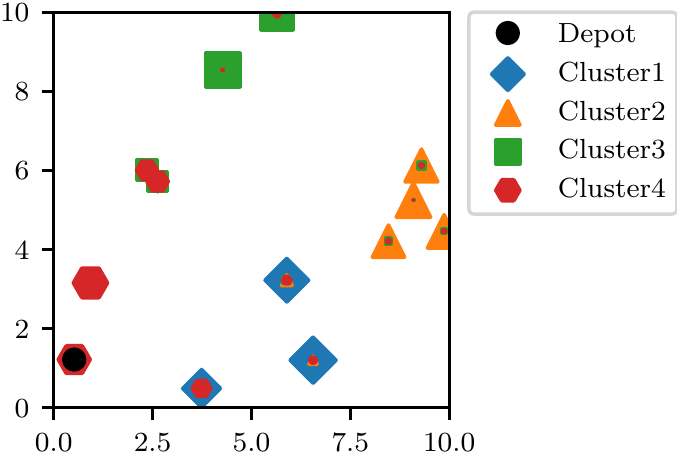} & 
\includegraphics[height=0.32\textwidth]{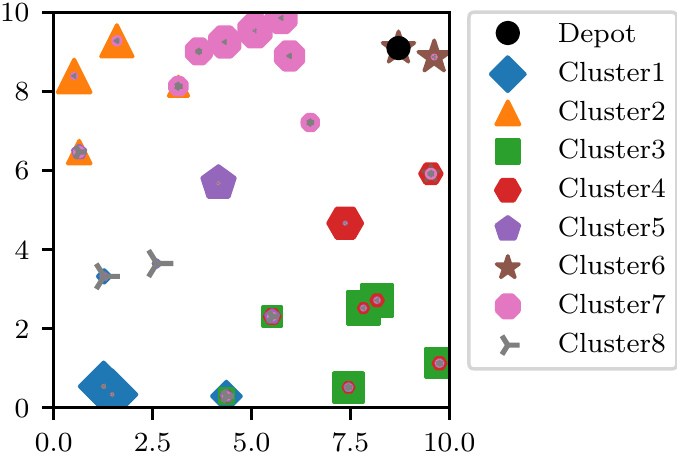}\\
(a) 4 robots and 12 tasks & (b) 8 robots and 24 tasks
\end{tabular}
\caption{Clusters given by FCM for the test cases 2 and 3 (Table \ref{tbl:resultTestScenarios}). The size of markers is proportional to the corresponding cluster coefficient of each task w.r.t. the associated cluster. \textit{The left plot shows that two tasks at point (2.5,6) are simultaneously classified into Clusters 3 and 4.}}
\label{fig:cse610hw1p3_cluster}
\end{figure*}
The prescribed parameter, $K_b$, is set at 1000. We define 7 different case studies, corresponding to different combinations of numbers of ground robots and tasks; allowed maximum tasks per robot is kept at $p=\lfloor n/m+2 \rceil$ to promote equitable distribution of load across the team. These test cases are listed in Table \ref{tbl:resultTestScenarios}. The task locations are randomly distributed in a $10\times 10$ sized 2D space. In order to provide a statistically insightful evaluation and comparison, ten random test scenarios are generated for each case. As the dimensionality of the robot-task-space increases, so does the computational cost of clustering. Hence, it is important to explore the impact and cost of clustering, or more specifically how the algorithm performs in the presence and in the absence of clustering. To this end, the following three different implementations of the decentralized MATA algorithm are tested:
\vspace{-0.2cm}
\begin{enumerate}
\item Dec-MATA: the original algorithm with clustering (Fig.~\ref{fig:cse610hw1p3_DecMata});\vspace{0.1cm}
\item DM-No-FCM-1.00: a modified algorithm where no FCM clustering is used, and the the clustering coefficients (used in bigraph construction) are replaced with a constant value of $b^*_{ij} = 1.00, \forall(i,j)\in(R,J)$; and \vspace{0.1cm}
\item DM-No-FCM-0.01: same as the previous implementation (i.e., no FCM clustering), except that the clustering coefficients are set at $b^*_{ij} = 0.01, \forall(i,j)\in(R,J)$.
\end{enumerate}

\begin{table*}[bph!]
\centering
\caption{Performance results of the algorithms on 7 test cases} \label{tbl:resultTestScenarios}
\footnotesize
\begin{tabular}{ccccccccc}
\toprule[0.12em]
\textbf{Case} & \textbf{\# of Tasks} & \textbf{\# of Agents} & \textbf{Max. Tasks per Robot} &  \multicolumn{4}{c}{\textbf{Overall Cost ( $\sigma(Agents' Cost)$ )}}\\
 & $n$ & $m$ & $h$ & Centralized & Dec-MATA & DM-No-FCM-1.00 & DM-No-FCM-0.01\\
\midrule[0.12em]
1 & 10 & 2 & 7 & 35.56 ( 3.74 )  &  41.65 ( 2.99 )  &  43.2 ( 2.21 )  &  43.36 ( 2.51 )\\
2 & 12 & 4 & 5 & 54.27 ( 4.73 )  &  64.81 ( 4.15 )  &  74.25 ( 2.65 )  &  74.68 ( 3.06 )\\
3 & 24 & 8 & 5 & 92.69 ( 5.26 )  &  113.25 ( 4.58 )  &  138.79 ( 3.04 )  &  139.57 ( 3.16 )\\
4 & 20 & 10 & 4 & 113.54 ( 4.96 )  &  121.6 ( 5.19 )  &  149.02 ( 2.94 )  &  148.03 ( 2.93 )\\
5 & 60 & 20 & 5 & 197.74 ( 5.33 )  &  254.91 ( 5.13 )  &  338.24 ( 3.1 )  &  338.09 ( 3.17 )\\
6 & 80 & 4 & 22 & 97.12 ( 4.27 )  &  122.81 ( 5.67 )  &  122.5 ( 3.69 )  &  122.39 ( 4.32 )\\
7 & 100 & 50 & 4 & 551.74 ( 4.53 )  &  593.77 ( 4.68 )  &  742.56 ( 3.2 )  &  740.18 ( 3.31 )\\
%8 & 100 & 300 &  626.75 (5.17) & \textbf{332.37} (2.49) & 415.66 (1.60) & 414.68 (1.63)\\
\bottomrule[0.12em]
\end{tabular}
\end{table*}

\vspace{-0.2cm}
\subsection{Results: Impact of FCM \& Bigraph Construction}\label{ssec:results}
Example illustrations showing outcomes of fuzzy clustering and bigraph construction are given first to provide insights on the impact of these steps of Dec-MATA. For example, Fig.~\ref{fig:cse610hw1p3_cluster} illustrates how the FCM divides the tasks into a set of $m$ clusters ($m$: \# robots) for the test cases 2 and 3 (Table \ref{tbl:resultTestScenarios}). The size of the marker in these plots is proportional to the corresponding cluster coefficient of each task. Since FCM is a soft clustering method, some tasks have two or more colors -- for example, two tasks at point (2.5,6) in Fig.~\ref{fig:cse610hw1p3_cluster}~(a) belong to both cluster 3 and 4. Figure~\ref{fig:cse610hw1p3_wbg} shows how the bigraph changes after the first iteration during the application of Dec-MATA for the test case 2 (Table \ref{tbl:resultTestScenarios}). It is readily evident, from the observed variation in the edge weights across robot--task pairs, how the bigraph is able to capture and alleviate the complexity of the task-assignment problem.

\subsection{Case Study Results: Performance Analysis}\label{ssec:results}
\begin{figure*}[h!]
\centering
\begin{tabular}{cc}
\includegraphics[height=0.3\textwidth]{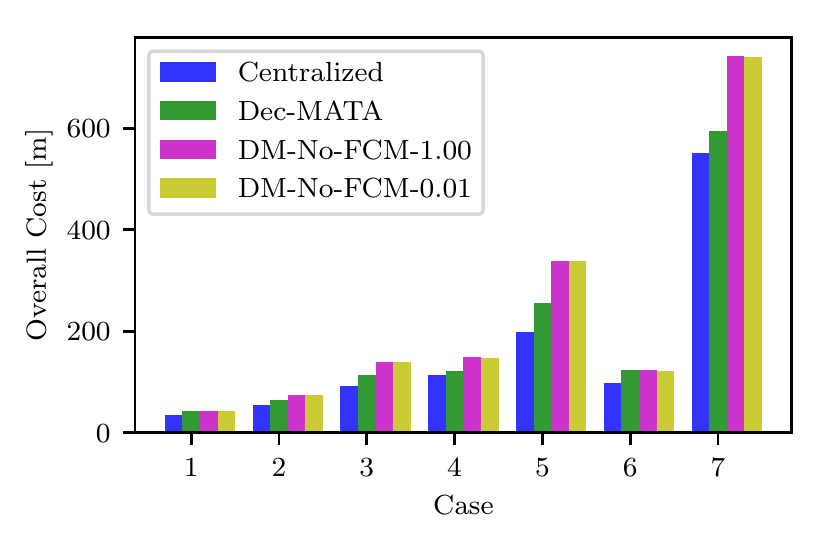} & 
\includegraphics[height=0.3\textwidth]{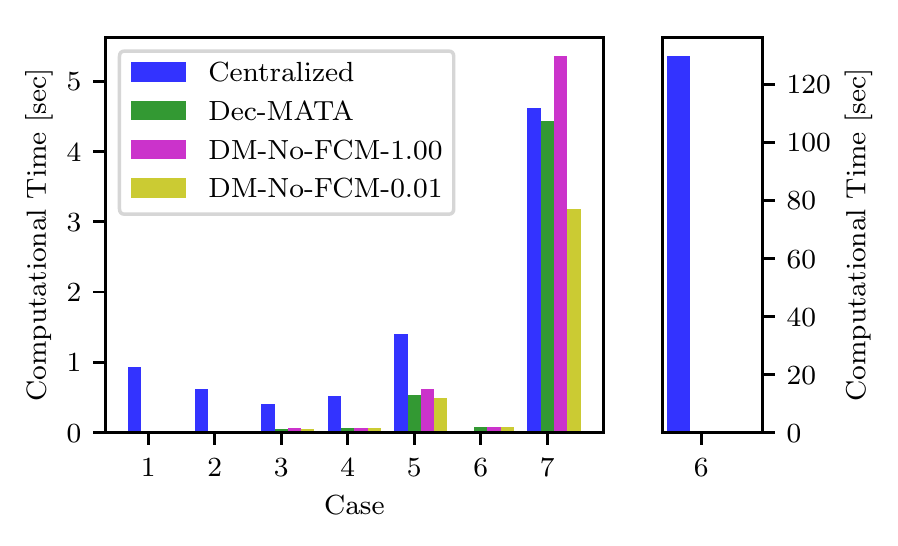}\\
(a) The best overall cost at termination of each algorithm & (b) Computing time w.r.t. terminal performance state of Dec-MATA
\end{tabular}
\caption{Performance and computing time of the algorithms for the 7  case studies. The computing time reported for the benchmark centralized algorithm is the approximate time required by it to reach the performance (cost) achieved by the Dec-MATA method; note that the centralized approach takes significantly (mostly several orders of magnitude) greater time to achieve its terminal optimal results depicted in (a).}
\label{fig:cse610hw1p3_scalability_rel}
\end{figure*}

Table~\ref{tbl:resultTestScenarios} summarizes the performances of the benchmark centralized mTSP/MILP algorithm and the three different Dec-MATA implementations, for the seven test cases. Both the \textbf{overall cost} (as given by Eq.~\eqref{eq:overallCost} incurred by the robot team under each algorithm, and the variance ($\sigma(Agents' Cost)$) in the individual costs incurred by the robots within the team are reported in the table. The reported values are estimated by averaging over the ten different runs of each case (with randomized depot and task locations). The overall cost incurred by each algorithm at their individual termination for each case is also shown as a bar plot in Fig. \ref{fig:cse610hw1p3_scalability_rel}(a). It can be seen from Table~\ref{tbl:resultTestScenarios} that the Dec-MATA solutions get promisingly close to the optimal solutions obtained by the benchmark centralized MILP algorithm (within 7\%--29
\% difference in overall cost). The observed variance in individual robot costs is comparable between the centralized and our Dec-MATA approach. The Dec-MATA approach also clearly outperforms the two other implementations (DM-no-FCM-1.00 and DM-no-FCM-0.01) where clustering is not used; this observation highlights the importance of clustering in the Dec-MATA algorithm. When a fixed value of $b_{ij}$ is used (across all robot--task pairs) in the absence of clustering, the value itself appears to have a negligible impact, given the (observed) practically similar performance of DM-no-FCM-1.00 and DM-no-FCM-0.01. 

\begin{figure}[h!]
\centering
\includegraphics[width=0.5\textwidth]{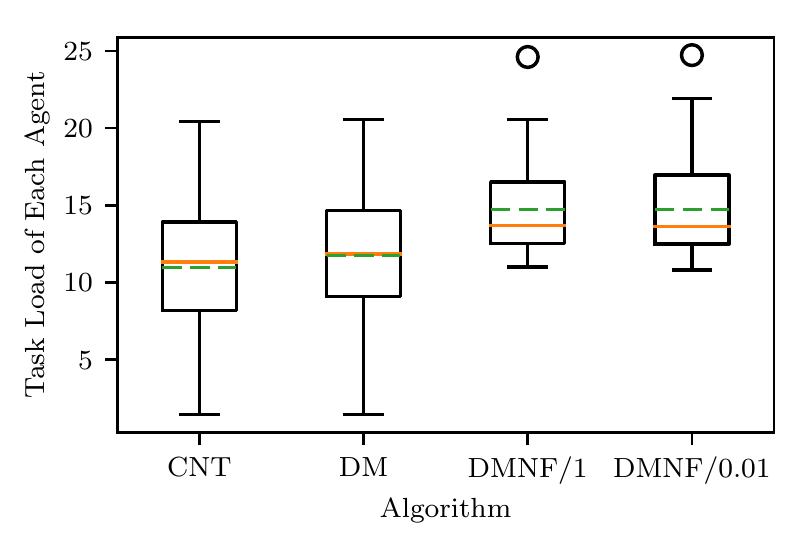}
\caption{Boxplots showing the variation in task load per robot across the team for test case 7 ($m=50$ and $n=100$), as accomplished by the following four algorithms: CNT: Centralized, DM: Dec-MATA, DMNF/1: DM-No-FCM-1.00, and DMNF/0.01: DM-No-FCM-0.01}
\label{fig:cse610hw1p3_BoxPlot}
\end{figure}

\begin{figure}[h!]
\centering
%\begin{tabular}{cc}
\includegraphics[width=0.5\textwidth]{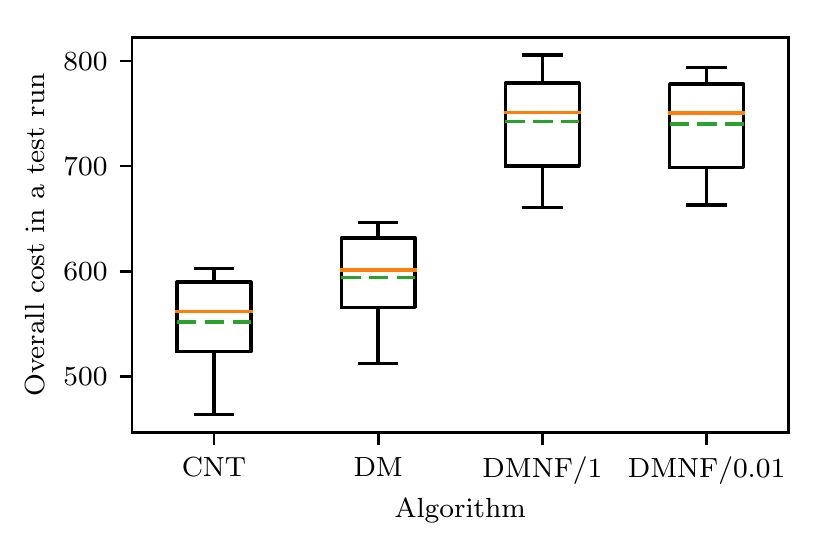} 
%\includegraphics[width=0.5\textwidth]{Figures/mata_milp_03-Jun-2018_n50_p4_j100_Run_DevPerTest.pdf}\\
%(a) & (b)   
%\end{tabular}
\caption{Boxplots showing the variation in overall cost across ten runs of test case 7 ($m=50$ and $n=100$), as accomplished by the following four algorithms: CNT: Centralized, DM: Dec-MATA, DMNF/1: DM-No-FCM-1.00, and DMNF/0.01: DM-No-FCM-0.01}
\label{fig:cse610hw1p3_tc6_10times}
\end{figure}

One of the most complex test cases, case 7 (with 50 agents and 100 tasks) is chosen for further analysis. Figure~\ref{fig:cse610hw1p3_BoxPlot} shows the across-the-team variation in the cost incurred or task load experienced by individual robots. While the task load variation obtained by Dec-MATA is very similar to that obtained by the benchmark centralized MILP algorithm, it is interesting to note that the overall variance in individual robot task load reduces in the absence of clustering. In order to judge the robustness of Dec-MATA in comparison to the centralized algorithm, Fig.~\ref{fig:cse610hw1p3_tc6_10times} illustrates the boxplot of the overall costs incurred by each algorithm across the ten runs of case 7. Variation in performance is expected, since the depot and task locations are randomized across the ten runs; what is interesting to note is that the overall variance in performance of the Dec-MATA algorithm is comparable to that of the centralized algorithm, thereby highlighting the robustness of Dec-MATA. %and compare the robustness of the discussed algorithms, all case studies are run for 10 times (with randomized generation of task locations in the 2D space each time) and the overall cost is presented in Fig.~\ref{fig:cse610hw1p3_tc6_10times} for the case 7. In terms of the overall cost, the centralized and Dec-MATA algorithm show the best performance. 
%In terms of the variation, the DM-No Clustering and Dec-MATA demonstrate slightly better performance.
%
\begin{table*}[bph!]
\centering
\caption{Computation times of Dec-MATA and mTSP on 7 test cases. $t_{dm}$: total computation time of Dec-MATA; $t_\text{reach}$: computation time of centralized MILP to reach Dec-MATA's performance; and $t_\text{opt}$: computation time of centralized MILP to reach the final/optimal solutions.} \label{tbl:resultTestTime}
\footnotesize
\begin{tabular}{cccccccc}
\toprule[0.12em]
\textbf{Case} & \textbf{\# of Tasks} & \textbf{\# of Agents} & \textbf{Max. Tasks per Robot} &  \multicolumn{3}{c}{\textbf{Computation Time} [sec]}\\
 & $n$ & $m$ & $h$ & $t_{dm}$ & $t_\text{reach}$ & $t_\text{opt}$\\
\midrule[0.12em]
1 & 10 & 2 & 7 & 0.01  &  0.94  &  1.18\\
2 & 12 & 4 & 5 & 0.02  &  0.63  &  0.82\\
3 & 24 & 8 & 5 & 0.06  &  0.41  &  7.67\\
4 & 20 & 10 & 4 & 0.06  &  0.52  &  0.65\\
5 & 60 & 20 & 5 & 0.53  &  1.4  &  6851.38\\
6 & 80 & 4 & 22 & 0.08  &  129.74  &  6004.65\\
7 & 100 & 50 & 4 & 4.44  &  4.62  &  4.63\\
%8 & 100 & 300 &  626.75 (5.17) & \textbf{332.37} (2.49) & 415.66 (1.60) & 414.68 (1.63)\\
\bottomrule[0.12em]
\end{tabular}
\end{table*}
%
% \begin{table*}[bpht]
% \centering
% \caption{The results of the test scenarios.} \label{tbl:resultTestScenarios}
% \footnotesize
% \begin{tabular}{lcccccccc}
% \toprule[0.12em]
% Case & \# of Agents & \# of Tasks & Max. Tasks per Robot &  \multicolumn{4}{c}{Overall Cost ($\sigma(Agents' Cost)$)}\\
%  & $m$ & $n$ & $h$ & Centralized & Dec-MATA & DM-No Clustering & DM-Fixed Weight\\
% \midrule[0.12em]
% 1 & 2 & 10 & 7 & \textbf{32.55} (2.53)  &  \textbf{41.65} (2.99)  &  43.28 (\textbf{2.33})  &  43.32 (2.36)\\
% 2 & 4 & 12 & 5 & \textbf{44.71} (2.79)  &  \textbf{64.81} (4.15)  &  74.25 (\textbf{2.65})  &  74.68 (3.06)\\
% 3 & 8 & 24 & 5 & \textbf{73.35} (\textbf{3.11})  &  \textbf{111.04} (4.75)  &  132.19 (3.89)  &  132.83 (3.99)\\
% 4 & 10 & 20 & 4 & \textbf{113.54} (\textbf{4.96})  &  \textbf{118.2} (\textbf{5.1})  &  128.1 (5.53)  &  128.73 (5.66)\\
% 5 & 20 & 60 & 5 & \\
% 6 & 50 & 100 & 4 & \\
% %8 & 100 & 300 &  626.75 (5.17) & \textbf{332.37} (2.49) & 415.66 (1.60) & 414.68 (1.63)\\
% \bottomrule[0.12em]
% \end{tabular}
% \end{table*}
%

%~\ref{fig:cse610hw1p3_tc6_10times} ~\ref{fig:cse610hw1p3_cluster}
%

\subsection{Case Study Results: Computational Efficiency}\label{ssec:results}
The comparison of the computational efficiency of Dec-MATA to that of the centralized algorithm is critical to analyzing (and thus appreciating) the advantages of decentralized schemes in multi-robot operations. Since the centralized algorithm and Dec-MATA are executed on two different platforms (MATLAB and the Jupyter/Python environment, respectively), a direct comparison is practically challenging. Thus, to obtain an understanding of how the two environments natively compare to each other, a basic \textit{for-loop} of length 1000 is executed 10 times on each platform, and the recorded computation is averaged across the 10 runs. This experiment showed that the MATLAB environment (that exploits all 8 CPU cores) is approximately 7.5 times faster than the Jupyter/Python environment. Thereafter, by directly (without any adjustment) comparing the recorded computation time of the centralized algorithm (run on the MATLAB environment) and the Dec-MATA algorithm (run on the Jupyter environment), the computation time advantages of Dec-MATA discussed below can be perceived as conservative.
%In order to obtain a reference point to compare the computation times in two different platforms, we run 10 times a for-loop with length 1000 and average the execution duration over 10 times. MATLAB  and Python (Jupyter environment) took 2 ms and 14.97 ms, respectively, which means the MATLAB is ~7.5 times faster than the Jupyter Python; it is reasonable as the MATLAB can utilize all eight available processors.
%
\begin{figure*}[h!]
\centering
\begin{tabular}{cc}
\includegraphics[height=0.3\textwidth]{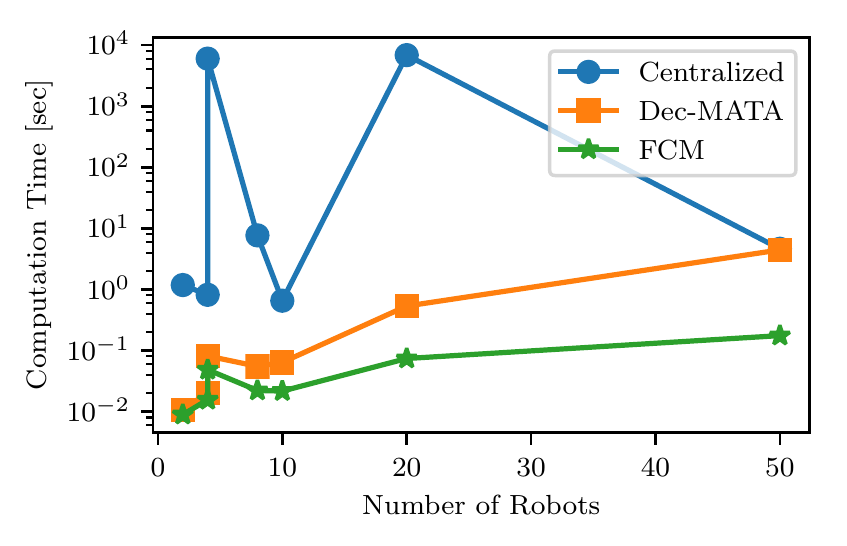} & 
\includegraphics[height=0.3\textwidth]{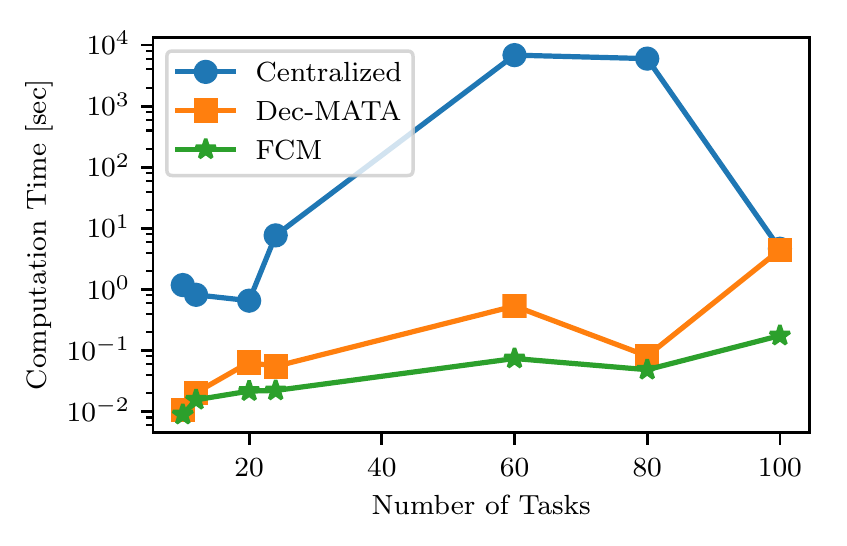}\\
(a) Computation time versus the number of robots & (b) Computation time versus the number of tasks
\end{tabular}
\caption{Variation of the computation time (till termination) of the Centralized, Dec-MATA, and the Dec-MATA's component FCM algorithms w.r.t. problem size. \textit{Dec-MATA's computation time is approximately 1-4 orders of magnitude smaller than that of the centralized algorithm at termination; while Dec-MATA's computation cost increases with problem size, it is minimally sensitive to the task-to-robot ratio unlike the centralized algorithm (which yields very high computation time for the 4 robot / 80 task case).}}
\label{fig:cse610hw1p3_scalability}
\end{figure*}

The following three different computation times are measured and analyzed: 1) total computation time of the Dec-MATA algorithm, termed $t_\text{dm}$; 2) computation time of the centralized MILP algorithm to reach/overtake Dec-MATA's performance, termed $t_\text{reach}$; and 3) total computation time of the centralized MILP algorithm to reach the final/optimal solutions ($t_\text{opt}$). %In order to interpret the obtained results, we define two important parameters: 
% \begin{enumerate}
% \item \textbf{Problem size:} the summation of the number of robots and tasks; i.e., $m+n$.
% \item \textbf{Task-robot ratio:} the ratio of the number of tasks to the number of robots; i.e., $n/m$.
% \end{enumerate}
These computation times for the 7 test cases (averaged across 10 randomized missions each) are summarized in Table~\ref{tbl:resultTestTime}. The computation times of Dec-MATA and the centralized algorithm at Dec-MATA's terminal performance level (i.e., $t_{dm}$ and $t_{reach}$) are also illustrated as bar plots in Fig. \ref{fig:cse610hw1p3_scalability_rel}~(b). For further analysis of how the problem size and characteristics impact the algorithms' computational efficiency, Figs. \ref{fig:cse610hw1p3_scalability}~(a)~and~(b) illustrate the variation in the algorithms' computation times with number of robots and number of tasks, respectively. 

It can be observed from Table \ref{tbl:resultTestTime} and Fig. \ref{fig:cse610hw1p3_scalability_rel}~(b) that the centralized algorithm, in general, required 1-3 orders of magnitude more computation time to reach the performance level of Dec-MATA. The total time required by the centralized algorithm to obtain its final/optimal is also fairly consistent w.r.t. to the time when it overtakes Dec-MATA's performance, except in the cases with a high tasks-to-robot ratio. This observation is also evident from Figs. \ref{fig:cse610hw1p3_scalability}~(a)~and~(b). Table \ref{tbl:resultTestTime} and \ref{fig:cse610hw1p3_scalability} show that while the cost of Dec-MATA scales sub-linearly with the problem size, it is minimally sensitive to the task-to-robot ratio. On the other hand, for the high task-to-robot ratio, the cost of the centralized algorithm spikes undesirably, e.g., it takes 129 seconds to reach Dec-MATA's performance level and 104 minutes to reach its optimal solution for the 4 robot / 80 task case. Lastly, Fig. \ref{fig:cse610hw1p3_scalability} shows that although the computation cost of the fuzzy clustering scales with the problem size, it always remains a small fraction of the total computation cost of Dec-DATA thereby justifying its role as an efficient intermediate approach to reduce the problem complexity of MATA. 

Overall, the sub-second computation time (as low as 10ms for the 2 robot / 10 task case) of Dec-MATA, except for the 100 robot case, and its negligible sensitivity to task-to-robot ratio clearly demonstrates Dec-MATA's promise in serving as a computationally tractable solution to autonomous multi-robot planning. Moreover, there remains an opportunity to further advance the computational efficiency of Dec-MATA by adopting more efficient (bipartite) maximum weighted matching implementations.

\section{CONCLUSION} \label{sec:conclusion}
In this paper, we developed a three-stage algorithm to perform decentralized allocation of location-based tasks in autonomous multi-robot systems, e.g., those catering to warehouse robotics and last-mile delivery applications. The decentralized multi-agent task-allocation or Dec-MATA algorithm aims to tackle the issue of exploding computational cost with increasing problem size (i.e., number of robots/tasks in the environment), in order to preserve tractability of use in online planning. To this end, the first and second stages of the algorithm respectively seek to reduce the dimensionality of the problem (to be solved by individual agents) and generate a compact representation of task-to-robot relevance; and the third stage of the algorithm allows each agent to identify optimal task sequences to undertake. A novel combination of fuzzy clustering, and bipartite graph construction and maximum weighted matching is investigated to design this three-stage Dec-MATA algorithm. 

To evaluate the performance of the new Dec-MATA algorithm, a benchmark centralized algorithm is formulated based on a state-of-the-art MILP solution of the multi-traveling salesman (mTSP) problem. Both algorithms are tested on a suite of case studies involving a varying number of robots and tasks (2 robot/10 task to 50 robot/100 task scenarios), and performance is analyzed in terms of the total cost incurred by the robot team (total distance traveled in this case) and computation time. While getting within 7-28\% of the minimum cost given by the centralized MILP algorithm, \textit{dec-MATA is observed to provide 1-3 orders of magnitude gain in computational efficiency (at comparable performance level), relatively similar robustness across randomized missions, and clearly superior scalability with increasing task-to-robot ratio}, when compared to the centralized algorithm.  %Although, the Dec-MATA provided sub-optimal solutions in comparison to the centralized approach in terms of performance, but it performed 2 to 240 times faster than the centralized implementation by obtaining a comparable solution.
%(by obtaining an {\color{red}{optimal}} solution with 20\% to 80\% less cost than the centralized implementation)
Dec-Mata solves all the test cases within a fraction of a second, except the largest case (50 robot/100 task), which requires 4.44 seconds. The latter observation indicates opportunities to further optimize the Dec-MATA algorithm by adopting more efficient problem decomposition, graph matching, and program parallelization approaches; these are the immediate next steps in this research. Moreover, the current version of Dec-MATA makes certain simplifying assumptions, which are likely reasonable for warehouse robotics-type applications. For example, decisions are synchronized across the team (no active consensus is however required), the task feature and agent states are assumed to be  precisely known, and each agent has a full observation of all tasks and the state of other agents. Future work will focus on alleviating these assumptions, thereby allowing wider application of Dec-MATA to asynchronous multi-robot problems involving uncertain environment and partial observability. %Future application of this framework on a wide variety of practical multi-robot problems would further establish its potential. 

%\newpage
%%%%%%%%%%%%%%%%%%%%%%%%%%%%%%%%%%%%%%%%%%%%%%%%%%%%%%%%%%%%%%%%%%%%%%
%\bibliographystyle{asmems4}
\bibliographystyle{IEEEtran}
\bibliography{payam2018map}

\begin{thebibliography}{10}
\providecommand{\url}[1]{#1}
\csname url@samestyle\endcsname
\providecommand{\newblock}{\relax}
\providecommand{\bibinfo}[2]{#2}
\providecommand{\BIBentrySTDinterwordspacing}{\spaceskip=0pt\relax}
\providecommand{\BIBentryALTinterwordstretchfactor}{4}
\providecommand{\BIBentryALTinterwordspacing}{\spaceskip=\fontdimen2\font plus
\BIBentryALTinterwordstretchfactor\fontdimen3\font minus
  \fontdimen4\font\relax}
\providecommand{\BIBforeignlanguage}[2]{{%
\expandafter\ifx\csname l@#1\endcsname\relax
\typeout{** WARNING: IEEEtran.bst: No hyphenation pattern has been}%
\typeout{** loaded for the language `#1'. Using the pattern for}%
\typeout{** the default language instead.}%
\else
\language=\csname l@#1\endcsname
\fi
#2}}
\providecommand{\BIBdecl}{\relax}
\BIBdecl

\bibitem{kuhn1992multi}
N.~Kuhn and J.~M{\"u}ller, ``Multi-agent planning,'' \emph{Deutsches
  Forschungszentrum f{\"u}r K{\"u}nstliche Intelligenz}, p.~61, 1992.

\bibitem{korsah2013comprehensive}
G.~A. Korsah, A.~Stentz, and M.~B. Dias, ``A comprehensive taxonomy for
  multi-robot task allocation,'' \emph{The International Journal of Robotics
  Research}, vol.~32, no.~12, pp. 1495--1512, 2013.

\bibitem{colistra2014problem}
G.~Colistra, V.~Pilloni, and L.~Atzori, ``The problem of task allocation in the
  internet of things and the consensus-based approach,'' \emph{Computer
  Networks}, vol.~73, pp. 98--111, 2014.

\bibitem{kalmar2017multiagent}
T.~Kalm{\'a}r-Nagy, G.~Giardini, and B.~D. Bak, ``The multiagent planning
  problem,'' \emph{Complexity}, vol. 2017, 2017.

\bibitem{bektas2006multiple}
T.~Bektas, ``The multiple traveling salesman problem: an overview of
  formulations and solution procedures,'' \emph{Omega}, vol.~34, no.~3, pp.
  209--219, 2006.

\bibitem{gutin2006traveling}
G.~Gutin and A.~P. Punnen, \emph{The traveling salesman problem and its
  variations}.\hskip 1em plus 0.5em minus 0.4em\relax Springer Science \&
  Business Media, 2006, vol.~12.

\bibitem{johnson1997traveling}
D.~S. Johnson and L.~A. McGeoch, ``The traveling salesman problem: A case study
  in local optimization,'' \emph{Local search in combinatorial optimization},
  vol.~1, pp. 215--310, 1997.

\bibitem{laporte1980cutting}
G.~Laporte and Y.~Nobert, ``A cutting planes algorithm for the m-salesmen
  problem,'' \emph{Journal of the Operational Research society}, vol.~31,
  no.~11, pp. 1017--1023, 1980.

\bibitem{gavish1986optimal}
B.~Gavish and K.~Srikanth, ``An optimal solution method for large-scale
  multiple traveling salesmen problems,'' \emph{Operations Research}, vol.~34,
  no.~5, pp. 698--717, 1986.

\bibitem{gromicho1992exact}
J.~Gromicho, J.~Paix{\~a}o, and I.~Bronco, ``Exact solution of multiple
  traveling salesman problems,'' in \emph{Combinatorial Optimization}.\hskip
  1em plus 0.5em minus 0.4em\relax Springer, 1992, pp. 291--292.

\bibitem{zhang1999team}
T.~Zhang, W.~Gruver, and M.~H. Smith, ``Team scheduling by genetic search,'' in
  \emph{Intelligent Processing and Manufacturing of Materials, 1999. IPMM'99.
  Proceedings of the Second International Conference on}, vol.~2.\hskip 1em
  plus 0.5em minus 0.4em\relax IEEE, 1999, pp. 839--844.

\bibitem{ryan1998reactive}
J.~L. Ryan, T.~G. Bailey, J.~T. Moore, and W.~B. Carlton, ``Reactive tabu
  search in unmanned aerial reconnaissance simulations,'' in \emph{Proceedings
  of the 30th conference on Winter simulation}.\hskip 1em plus 0.5em minus
  0.4em\relax IEEE Computer Society Press, 1998, pp. 873--880.

\bibitem{gorenstein1970printing}
S.~Gorenstein, ``Printing press scheduling for multi-edition periodicals,''
  \emph{Management Science}, vol.~16, no.~6, pp. B--373, 1970.

\bibitem{jonker1988improved}
R.~Jonker and T.~Volgenant, ``An improved transformation of the symmetric
  multiple traveling salesman problem,'' \emph{Operations Research}, vol.~36,
  no.~1, pp. 163--167, 1988.

\bibitem{nallusamy2009optimization}
R.~Nallusamy, K.~Duraiswamy, R.~Dhanalaksmi, and P.~Parthiban, ``Optimization
  of non-linear multiple traveling salesman problem using k-means clustering,
  shrink wrap algorithm and meta-heuristics,'' \emph{International Journal of
  Nonlinear Science}, vol.~8, no.~4, pp. 480--487, 2009.

\bibitem{khamis2015multi}
A.~Khamis, A.~Hussein, and A.~Elmogy, ``Multi-robot task allocation: A review
  of the state-of-the-art,'' in \emph{Cooperative Robots and Sensor Networks
  2015}.\hskip 1em plus 0.5em minus 0.4em\relax Springer, 2015, pp. 31--51.

\bibitem{shoham2008multiagent}
Y.~Shoham and K.~Leyton-Brown, \emph{Multiagent systems: Algorithmic,
  game-theoretic, and logical foundations}.\hskip 1em plus 0.5em minus
  0.4em\relax Cambridge University Press, 2008.

\bibitem{dai2011multi}
B.~Dai and H.~Chen, ``A multi-agent and auction-based framework and approach
  for carrier collaboration,'' \emph{Logistics Research}, vol.~3, no. 2-3, pp.
  101--120, 2011.

\bibitem{choi2009consensus}
H.-L. Choi, L.~Brunet, and J.~P. How, ``Consensus-based decentralized auctions
  for robust task allocation,'' \emph{IEEE transactions on robotics}, vol.~25,
  no.~4, pp. 912--926, 2009.

\bibitem{dAndrea2012guest}
R.~D'Andrea, ``Guest editorial: A revolution in the warehouse: A retrospective
  on kiva systems and the grand challenges ahead,'' \emph{IEEE Transactions on
  Automation Science and Engineering}, vol.~9, no.~4, pp. 638--639, 2012.

\bibitem{wurman2008coordinating}
P.~R. Wurman, R.~D'Andrea, and M.~Mountz, ``Coordinating hundreds of
  cooperative, autonomous vehicles in warehouses,'' \emph{AI magazine},
  vol.~29, no.~1, p.~9, 2008.

\bibitem{miller1960integer}
C.~E. Miller, A.~W. Tucker, and R.~A. Zemlin, ``Integer programming formulation
  of traveling salesman problems,'' \emph{Journal of the ACM (JACM)}, vol.~7,
  no.~4, pp. 326--329, 1960.

\bibitem{bezdek1984fcm}
J.~C. Bezdek, R.~Ehrlich, and W.~Full, ``Fcm: The fuzzy c-means clustering
  algorithm,'' \emph{Computers \& Geosciences}, vol.~10, no. 2-3, pp. 191--203,
  1984.

\bibitem{bezdek1981objective}
J.~C. Bezdek, ``Objective function clustering,'' in \emph{Pattern recognition
  with fuzzy objective function algorithms}.\hskip 1em plus 0.5em minus
  0.4em\relax Springer, 1981, pp. 43--93.

\bibitem{slota2018graph}
G.~M. Slota, ``General graph matching,'' in \emph{Graph Theory, Spring
  2017}.\hskip 1em plus 0.5em minus 0.4em\relax Rensselaer Polytechnic
  Institute, 2017.

\bibitem{huang2007analyzing}
Z.~Huang, D.~D. Zeng, and H.~Chen, ``Analyzing consumer-product graphs:
  Empirical findings and applications in recommender systems,''
  \emph{Management science}, vol.~53, no.~7, pp. 1146--1164, 2007.

\bibitem{zhu2015measuring}
Z.~Zhu, J.~Su, and L.~Kong, ``Measuring influence in online social network
  based on the user-content bipartite graph,'' \emph{Computers in Human
  Behavior}, vol.~52, pp. 184--189, 2015.

\bibitem{asratian1998bipartite}
A.~S. Asratian, T.~M. Denley, and R.~H{\"a}ggkvist, \emph{Bipartite graphs and
  their applications}.\hskip 1em plus 0.5em minus 0.4em\relax Cambridge
  University Press, 1998, vol. 131.

\bibitem{lovasz2009matching}
L.~Lov{\'a}sz and M.~D. Plummer, \emph{Matching theory}.\hskip 1em plus 0.5em
  minus 0.4em\relax American Mathematical Soc., 2009, vol. 367.

\bibitem{galil1986efficient}
Z.~Galil, ``Efficient algorithms for finding maximum matching in graphs,''
  \emph{ACM Computing Surveys (CSUR)}, vol.~18, no.~1, pp. 23--38, 1986.

\bibitem{edmonds1965maximum}
J.~Edmonds, ``Maximum matching and a polyhedron with 0, 1-vertices,''
  \emph{Journal of Research of the National Bureau of Standards B}, vol.~69,
  no. 125-130, pp. 55--56, 1965.

\bibitem{zedadra2017multi}
O.~Zedadra, N.~Jouandeau, H.~Seridi, and G.~Fortino, ``Multi-agent foraging:
  state-of-the-art and research challenges,'' \emph{Complex Adaptive Systems
  Modeling}, vol.~5, no.~1, p.~3, 2017.

\bibitem{gurobi}
\BIBentryALTinterwordspacing
I.~Gurobi~Optimization, ``Gurobi optimizer reference manual,'' 2016. [Online].
  Available: \url{http://www.gurobi.com}
\BIBentrySTDinterwordspacing

\end{thebibliography}

\end{document}